\documentclass[twocolumn]{aastex62}

\shorttitle{Flaring in pre-merger binary neutron stars magnetospheres}
\shortauthors{E.R. Most \& A.A. Philippov}

\usepackage{amsmath}
\usepackage{amssymb}

\begin{document}

\title{Electromagnetic precursors to gravitational wave events: \\
\it Numerical simulations of flaring in pre-merger binary neutron star magnetospheres}

\correspondingauthor{Elias R. Most}
\email{emost@itp.uni-frankfurt.de}

\author{Elias R. Most}
\affiliation{Center for Computational Astrophysics, Flatiron Institute, 162
Fifth Avenue, New York, NY 10010, USA}
\affiliation{Institut f\"ur Theoretische Physik, Goethe Universit\"at
Frankfurt am Main, Germany}

\author{Alexander A. Philippov}
\affiliation{Center for Computational Astrophysics, Flatiron Institute, 162
Fifth Avenue, New York, NY 10010, USA}


e

\begin{abstract}
  The detection of gravitational waves from neutron star merger events has
  opened up a new field of multi-messenger astronomy linking gravitional
  waves events to short-gamma ray bursts and kilonova afterglows.
  A further - yet to be discovered - electromagnetic counterpart is
  precursor emission produced by the non-trivial interaction of the magnetospheres
  of the two neutron stars prior to merger.
  By performing special-relativistic force-free simulations of orbiting
  neutron stars we discuss the effect of different magnetic field orientations and
  show how the emission can be significantly enhanced by differential
  motion present in the binary, either due to stellar spins or misaligned stellar magnetospheres.
  We find that the build-up of twist in the
  magnetic flux tube connecting the two stars can lead to the repeated emission of
  powerful flares for a variety of orbital configurations. We also
  discuss potential coherent radio emission mechanisms in the flaring process.
\end{abstract}



\section{Introduction}
\label{sec:intro}
%
Possessing some of the highest densities and strongest magnetic fields in the universe neutron
stars are an ideal tool to study strong gravity, nuclear physics and
high-energy plasma physics alike. 
The recent multi-messenger observation of the neutron star merger GW170817
has demonstrated how the different observational channels (gravitational
waves, kilonova afterglow and short gamma-ray burst) can be used to
constrain the properties of neutron stars, e.g. their masses, radii and, in turn,
nuclear physics beyond saturation
\citep{Abbott2018b,Most2018,De2018,Raithel2019}.  While the observed
electromagnetic counterparts have all been emitted following the merger,
the presence of strong magnetic field configurations in radio pulsars
indicates that the magnetospheres of the two stars might interact non-trivially prior to
the merger. Even if the two neutron stars themselves would have inactive
magnetospheres due to the spin-down over their long lifetime, the interaction of the two
stars can reignite pair creation and establish a nearly force-free \citep{Goldreich:1969} magnetosphere filled with pair plasma at the time of merger \citep{Lyutikov2018}.
This scenario has been shown analytically to
dissipate electromagnetic energy either in terms of a unipolar inductor
scenario \citep{Lai2012,Piro2012}, where one of the stars has a higher
magnetic field than its companion and also in the case of comparable
magnetization \citep{Hansen2001,Lyutikov2018}. In
addition this scenario has also been invoked to drive
powerful fireballs \citep{Metzger2016} and Fast Radio Bursts (FRBs) \citep{Wang2016}.
Since the highly dynamical electromagnetic field configurations present in the inspiraling
binary are too involved to be studied using purely analytical approaches a
few numerical studies have been performed in order to study force-free
magnetospheric interactions, either in binary black hole mergers
\citep{Alic:2012,Palenzuela:2010a}, in
neutron star binaries \citep{Palenzuela2013a, Palenzuela2013b, Ponce2014}, in mixed binaries
\citep{Paschalidis2013} and in collapsing neutron stars \citep{Lehner2011,Palenzuela2013} (see also 
\citet{Most2017,Nathanail2017} for electrovacuum simulations).  While these studies have been performed
self-consistently in full general relativity they have not studied the main source
of energy dissipation in current sheets, which are important sources of broad-band electromagnetic emission.
In this Letter we show that if all of the magnetic field dynamics is fully resolved the interaction of the magnetic
fields in the binary can lead to the launching of powerful magnetic flares
similar to magnetars \citep{Kaspi2017} and coronal mass ejection in the Sun \citep{Forbes2000}, where magnetic
energy dissipation occurs in the main current sheet trailing the flare.

\section{Methods}
\label{sec:methods}

This work studies the emission of electromagnetic precursor prior to the merger
of a double neutron star system.  We model the neutron stars as spherical
conductors with a circumferential radius of $13\,\rm km$ and a spin axis
aligned with the orbital angular momentum. The neutron stars are equipped with
dipole magnetic fields having a magnetic field strength $B_0$ at the surface.
We solve the covariant equations of general-relativistic force-free
electrodynamics \citep{Palenzuela2013} using the newly developed \texttt{GReX}
code \citep{Most2020}. We decompose the four-dimensional metric $g_{\mu\nu}$
within a 3+1 split into ${\rm d}s^2 = \left( -\alpha^2 + \beta_k \beta^k
\right) {\rm d} t^2 + 2 \beta_k {\rm d} x^k {\mathrm d} t +\gamma_{ij} {\rm
d}x^i {\rm d} x^j$. Since, for simplicity, we only incorporate special
relativistic effects we adopt a co-rotating Minkowski frame $\alpha=1$,
$\gamma_{ij} = \delta_{ij}$ and $\beta_i = - \varepsilon_{ijk} \Omega^j x^k$,
where ${\bf \Omega} = (0,0,0,\omega)$ is the orbial angular momentum vector,
\citep[see also][]{Schiff39,Carrasco2020}.  The interior of the neutron star
obeys the ideal MHD condition \begin{align} E^i = - \varepsilon^{ijk} v_j B_k,
\label{eqn:ideal} \end{align} where $v_i = \varepsilon_{ijk} \Omega_s^j x_s^k$,
$\Omega_s$ being the spin vector of the neutron star and $x_s$ being the
coordinate vector centered on the neutron star. We emphasize that adopting a
co-rotating frame allows us to cleanly separate the orbital motion from the
spin of the individual neutron stars so that only the spin velocity enters in
\eqref{eqn:ideal}.  While simulations of pulsar magnetospheres typically do not
adopt a fully general-relativistic framework to solve the Maxwell equations in
the corotating frame  \citep[see e.g.][]{Bai2010}, 
it would be straightforward and interesting to also implement this approach
within relativisitc particle-in-cell codes for magnetospheric modeling \citep{Parfrey2019,Crinquand2020}.
More details can be found in the Appendix.  The
exterior is evolved according to the force-free conditions $E_i B^i =0$ and
$E^2 < B^2$ \citep{Komissarov2004b}.  These are then imposed using a stiff
constraint relaxation scheme \citep{Alic:2012} for the evolution of the
electric and magnetic fields $E^i$ and $B^i$ \citep{Baumgarte2003}. To handle
the stiff current in the Ohm's law we use the third order accurate
implicit-explicit (IMEX) Runge-Kutta(RK) SSP3(4,3,3) scheme
\citep{pareschi_2005_ier} and compute the explicit fluxes using a fourth order
accurate finite volume scheme \citep{mccorquodale2011high}, combining a
fifth-order WENO-Z reconstruction \citep{Borges2008} with a Rusanov Riemann
solver \citep{Rusanov1961a} similar to the approach in \citep{Most2019b}.  We
similarly enforce the ideal-MHD condition inside the star by computing a stiff
instantaneous correction current that exactly enforces \eqref{eqn:ideal} at
every substep of the IMEX-RK scheme.

The computational grid is provided by a set of nested boxes using the AMReX
\citep{Almgren2019} highly parallel adaptive mesh-refinement framework, on
which \texttt{GReX} is built. The outer-most box extends up to $\simeq
1200\,{\rm km}$ whereas the inner-most box spans $\simeq 40\, \rm km$ in all
directions with a highest resolution of $\simeq 400 \,{\rm m}$ and a total
number of $6$ refinement levels.  Since the angular component of the shift
$\beta_\phi = - \omega R$ diverges at large distances $R$ from the origin, we
damp the shift $\beta_\phi \simeq R^{-2}$ close to the boundary and impose
simple third order extrapolation boundary conditions on the electromagnetic
fields.
\begin{figure}[b]
  \centering
  \includegraphics[width=0.5\textwidth]{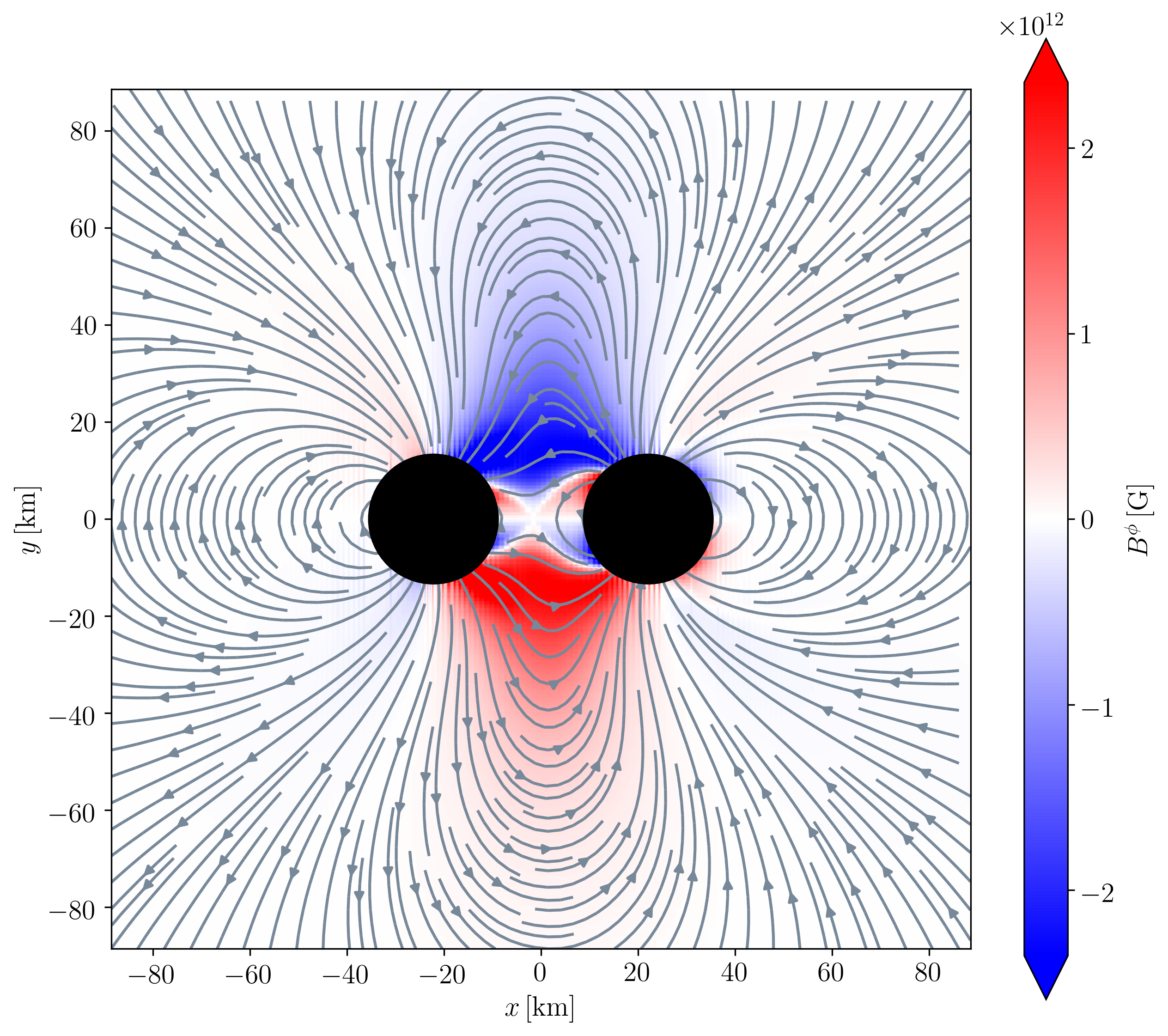}
  \caption{Intermediate force-free magnetic field configuration of a binary
  in close orbit, in which the right star is spinning. The color
  qualitatively indicates the twist, i.e. the out-of-plane 
  component $B^\phi$ of the magnetic field.}
  \label{fig:twist}
\end{figure}

\section{Results}
\label{sec:results}

\begin{figure}[t!]
  \centering
  {\includegraphics[width=0.5\textwidth]{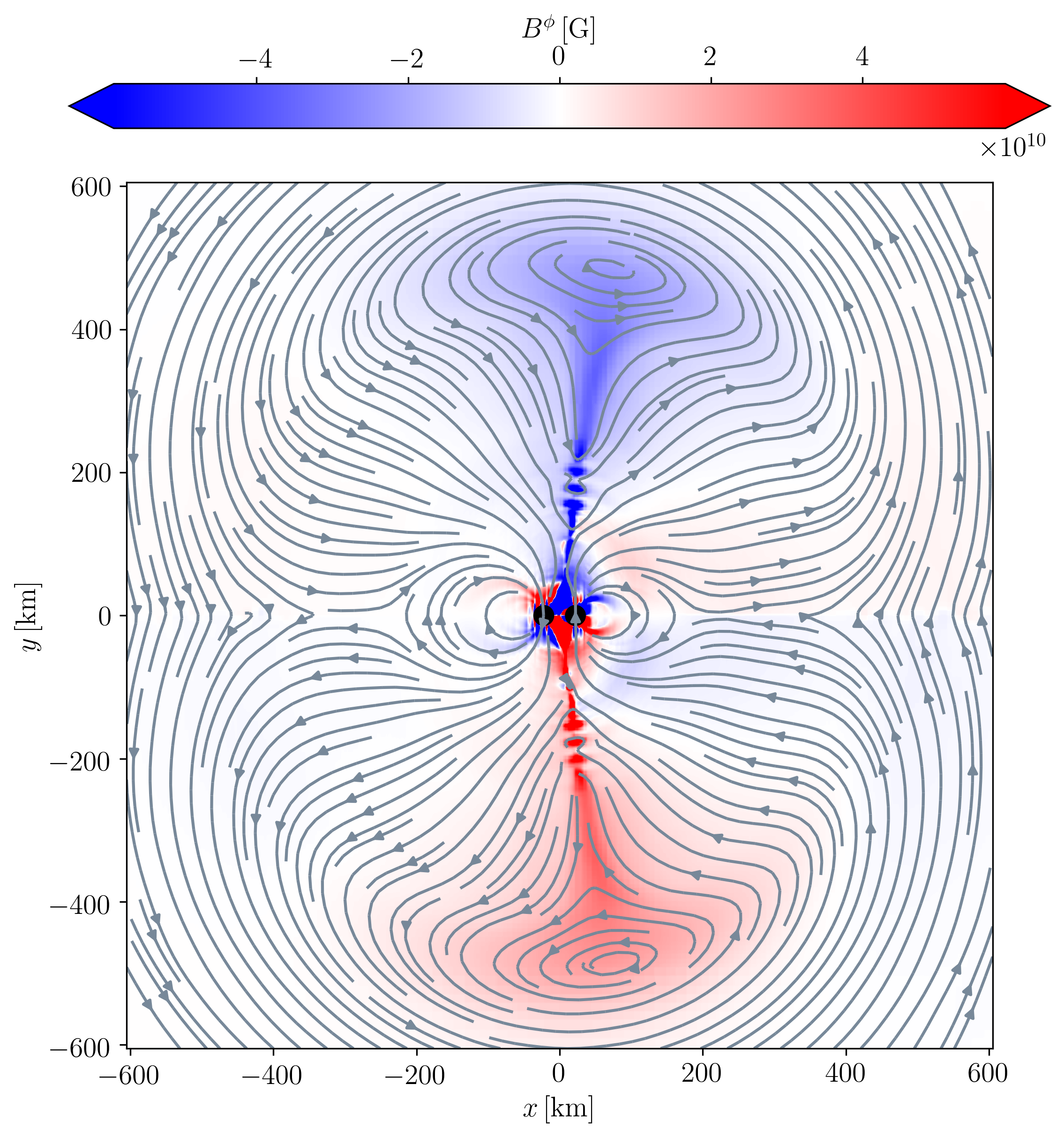}}
  \caption{Electromagnetic flare launched from an orbiting neutron star
  binary in close contact after a significant twist has built up due to the
  rotation of the right star. Shown in color is the out-of-plane magnetic field $B^\phi$ in
  a co-rotating frame indicating the twist in the flux tubes connecting the
  stars. The current sheet trailing the flare is shown to form magnetic
  islands as a result of the plasmoid instability. 
 }
  \label{fig:emission_al}
\end{figure}

We investigate the evolution of the common magnetosphere of a neutron star
binary in close orbit to demonstrate under which conditions powerful electromagnetic flares can be expected shortly before merger. We will first study a fiducial system in which
both stars are in a synchronised orbit, with one star having an additional spin $\Omega_s$,
and the energy for the flare is extracted
from the rotational energy of the spinning neutron star.
In order to establish that this flaring effect is also present in binaries
endowed with different field configurations we
in addition consider binaries where the magnetic moment of one of the stars
is misaligned with the orbital angular momentum. Finally, we also establish the robustness of our results by studying the fiducial binary at various separations ranging up to $100\, \rm km$. {As we show below, for the case of the misaligned binary the flaring occurs even in the case of non-spinning neutron stars. Moreover, the flaring luminosity depends only on the separation between the two stars, not on the actual values of the stellar spin. 
This proves that flaring events are driven by the energy stored in the twisted magnetic field loop, and not by the Poynting flux driven by the orbital or rotational motion \citep{Carrasco2019,Carrasco2020}.}

We consider two stars endowed with strong dipole
magnetic fields having a field strength $B_0 \simeq 10^{12}\, \rm G$ at the surface.
In particular, we focus on anti-aligned field line configurations with
their magnetic moments pointing in opposite directions, which result in a
closed loop magnetic field configuration. Since the field strengths of the
interacting dipoles become strongest at close separation shortly before
merger, we study a fiducial binary at $45\, \rm km$ separation, which corresponds to an orbital light cylinder $R_{LC}^{\rm orbit} \simeq 170\, \rm km$. The light cylinder here denotes the distance from the origin of the binary beyond which its no longer possible for the field to co-rotate as this would have to happen at speeds larger than the light speed \citep{Goldreich:1969}. We include differential motion in the fiducial binary by adding spin to one of the stars, corresponding to a stellar light cylinder $R_{LC}^{\rm star}\simeq 470\, \rm km$. As outlined in the previous section we then proceed
and perform a fully special relativistic force-free simulation of this
orbiting binary to illustrate how powerful electromagnetic flares can be
launched in a binary system in close contact prior to merger.


After an initial transient necessary for the force-free constraint
relaxation scheme to establish an initial force-free magnetic field
configuration, the spinning star in the fiducial binary continuously twists
the magnetic flux tube connecting the two stars. This is illustrated
in Fig. \ref{fig:twist} which shows the out-of-plane magnetic field
component $B^\phi$ in a co-rotating frame. The twist causes the magnetic field lines
to inflate transferring energy from the rotation of the neutron star into the
magnetic field. It is important that for this process to work the stars
need to be in sufficiently close contact since the twist is established by
an Alfven wave propagating between the two stars along the flux tube, which
requires that the separation $a$ of the binary is smaller than the stellar
light cylinder $R_{LC}^{\rm star}$. Assuming realistic dimensionless spins $\chi \leq 0.05$ \footnote{The dimensionless spin is defined as $\chi = J/M^2$, where $J$ and $M$ are the angular momentum and mass of the neutron star, respectively.}, i.e. $R_{LC}^{\rm star} \simeq 350 \rm km$,  in the binary
\citep{ZhuX2018}, this will always happen during the last orbits before the merger.

\begin{figure*}[t]
  \centering
  \includegraphics[width=0.45\textwidth]{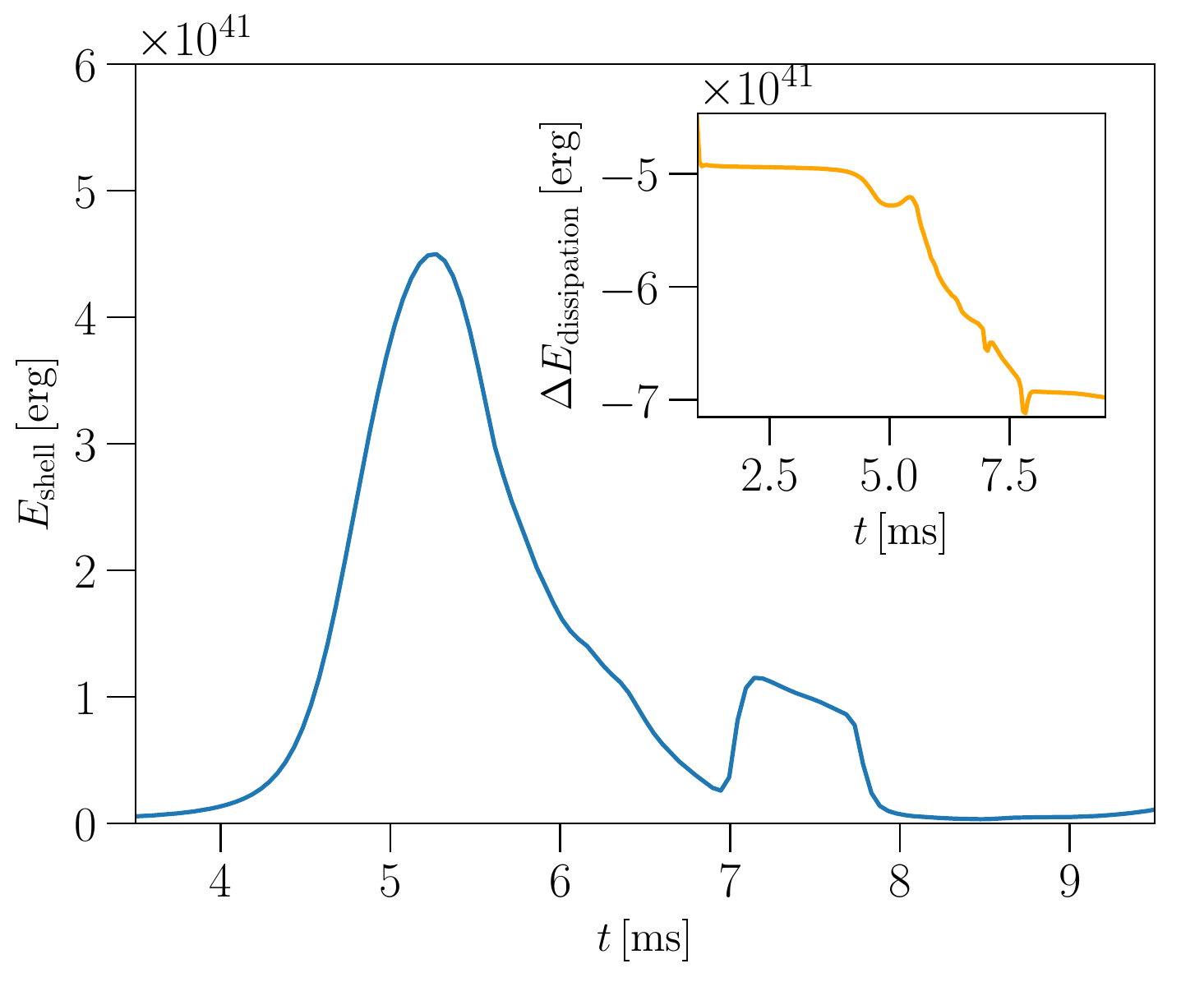}
  \includegraphics[width=0.47\textwidth]{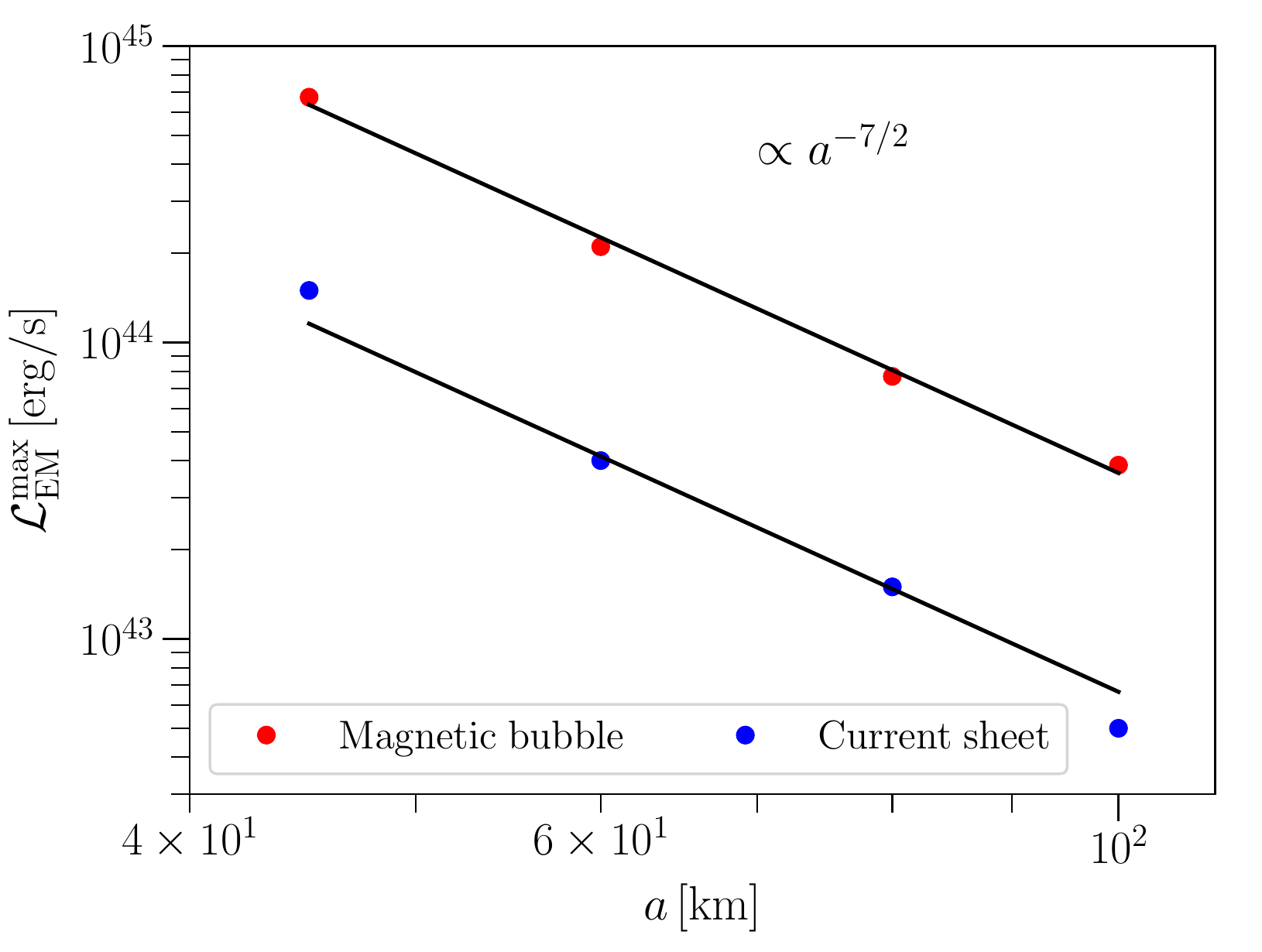}
  \caption{{(\it Left)}: Energy $E_{\rm shell}$ contained in the shell enclosing the flare. The inset is showing the energy $\Delta E_{\rm dissipation}$ dissipated in the current sheet. {(\it Right)}: Peak luminosity $\mathcal{L}_{\rm EM}^{\rm max}$ of the out-going magnetic bubble (red dots) and dissipation in the current sheet (blue dots) as a function of the separation $a$ between the stars in the binary.}
  \label{fig:en_dep}
\end{figure*}

As can already be anticipated in Fig. 1, at some point the built-up pressure of the toroidal magnetic field $B^\phi$ will be so strong that the twisted magnetic flux tube that connects the two stars, blue and red region in Fig. \ref{fig:twist}, has to open up.
This is shown in Fig. \ref{fig:emission_al} which presents the flare at the time when a magnetic bubble gets ejected together with a reconnecting current sheet trailing it, similar to magnetar flares (for example, in 2D force-free simulations of \citealt{Parfrey2012}). Looking at the out-of-plane magnetic field component it can easily be seen that magnetic islands are formed in the current sheet, suggesting that magnetic reconnection and the plasmoid instability are taking place 
\citep{Loureiro2007,Bhattacharjee2009}.
\begin{figure}[b]
  \centering
  \includegraphics[width=0.5\textwidth]{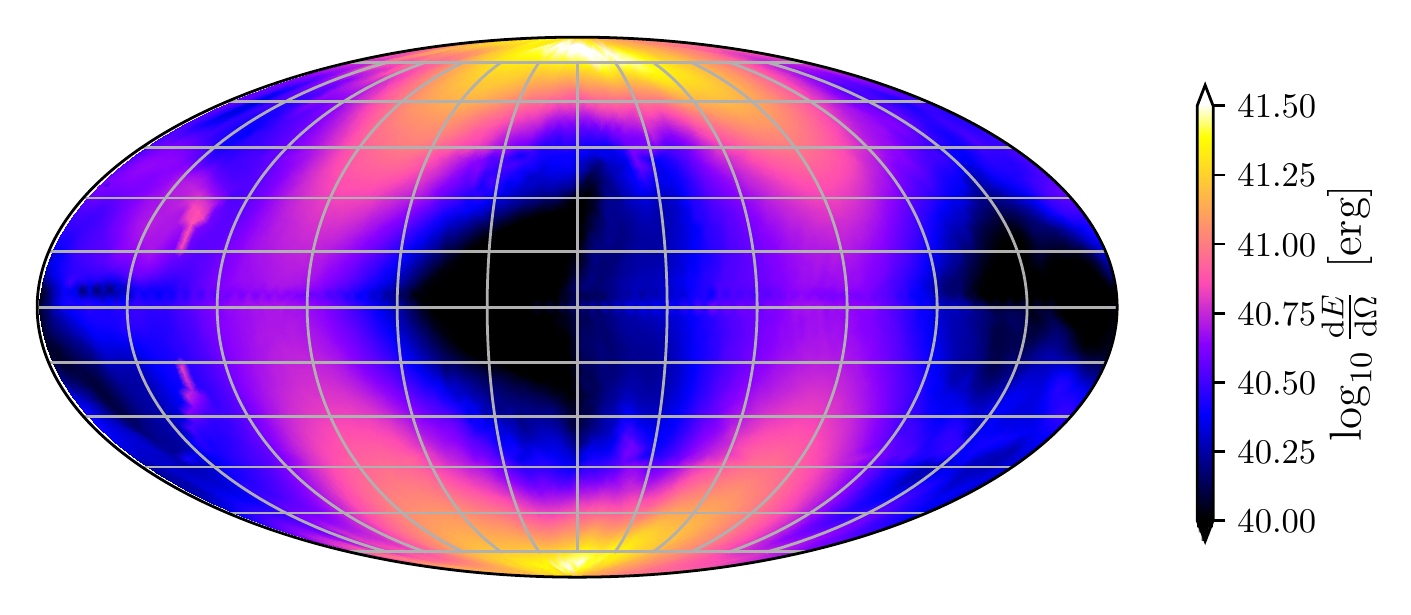}
  \caption{Time integrated electromagnetic energy ${\rm d} E/{\rm d}\Omega$  per surface angle
  extracted on a radial shell at $450\, \rm km$ radius.}
  \label{fig:poynting2D}
\end{figure}
In order to better quantify the energy contained in the out-going bubble as
well as understand the angular distribution of the emission on large
scales we extract the Poynting flux $4\pi {\bf S_{\rm EM} = E \times B} - \frac{1}{2} \left(E^2 +B^2\right){\boldsymbol \beta}$ on a spherical shell placed at a radius of $450\, \rm km$ from the origin. We then integrate the out-going energy flux over one
burst and show its angular distribution in Fig. \ref{fig:poynting2D}.
We can see that the bubble begins to widen at large scales from the binary
indicating its quasi-isotropic structure at infinity. Further, the energy in the bubble is about $10^{41}\, \rm erg$ per surface angle.
In order to quantify the amount of energy dissipated in the current sheet we consider Poynting's theorem of electromagnetic energy conservation in the simulation domain
\begin{align}
  \partial_t \left( \frac{1}{8\pi}\left( E^2 +B^2 \right) \right) + \nabla
  \cdot {\bf S_{\rm EM}} = \mathcal{L}_{\rm dissipation},
  \label{eqn:poynting}
\end{align}
where $\mathcal{L}_{\rm dissipation}$ is the dissipative luminosity, driven by the resistive terms in the Ohms law. Eq. \eqref{eqn:poynting} states the the electromagnetic energy $E_{\rm EM} = \int {\rm d}^3 x \frac{1}{8\pi}\left( E^2 +B^2 \right)$ can only change
either by a transport of energy via a Poynting flux ${\bf S}_{\rm EM}$ or via
dissipation $\mathcal{L}_{\rm dissipation}$. In order to estimate the amount of dissipation we consider two spherical surfaces centered on the origin of the binary.
The inner surface encloses just the two stars while the outer shell is placed
at a large distance from the origin. We can then estimate the amount of
dissipation by computing the energy balance in the shell between the two surfaces 
according to Eq. \eqref{eqn:poynting}. This is shown in the left panel of Fig. \ref{fig:en_dep} which shows the evolution of the energy $E_{\rm shell}$ contained in the shell during one flaring event and the amount of energy dissipation $\Delta E_{\rm dissipation}$ in the current sheet. We anticipate these results not to depend strongly on numerical resolution as it has been shown that the non-linear development of the plasmoid instability causes the magnetic reconnection rate to become independent of the physical resistivity $\eta$ at high Lundquist number $S_L \geq \eta^{-1} \simeq 10^4$ \citep{Loureiro2007,Bhattacharjee2009}.

\begin{figure}[t]
  \centering
  \includegraphics[width=0.47\textwidth]{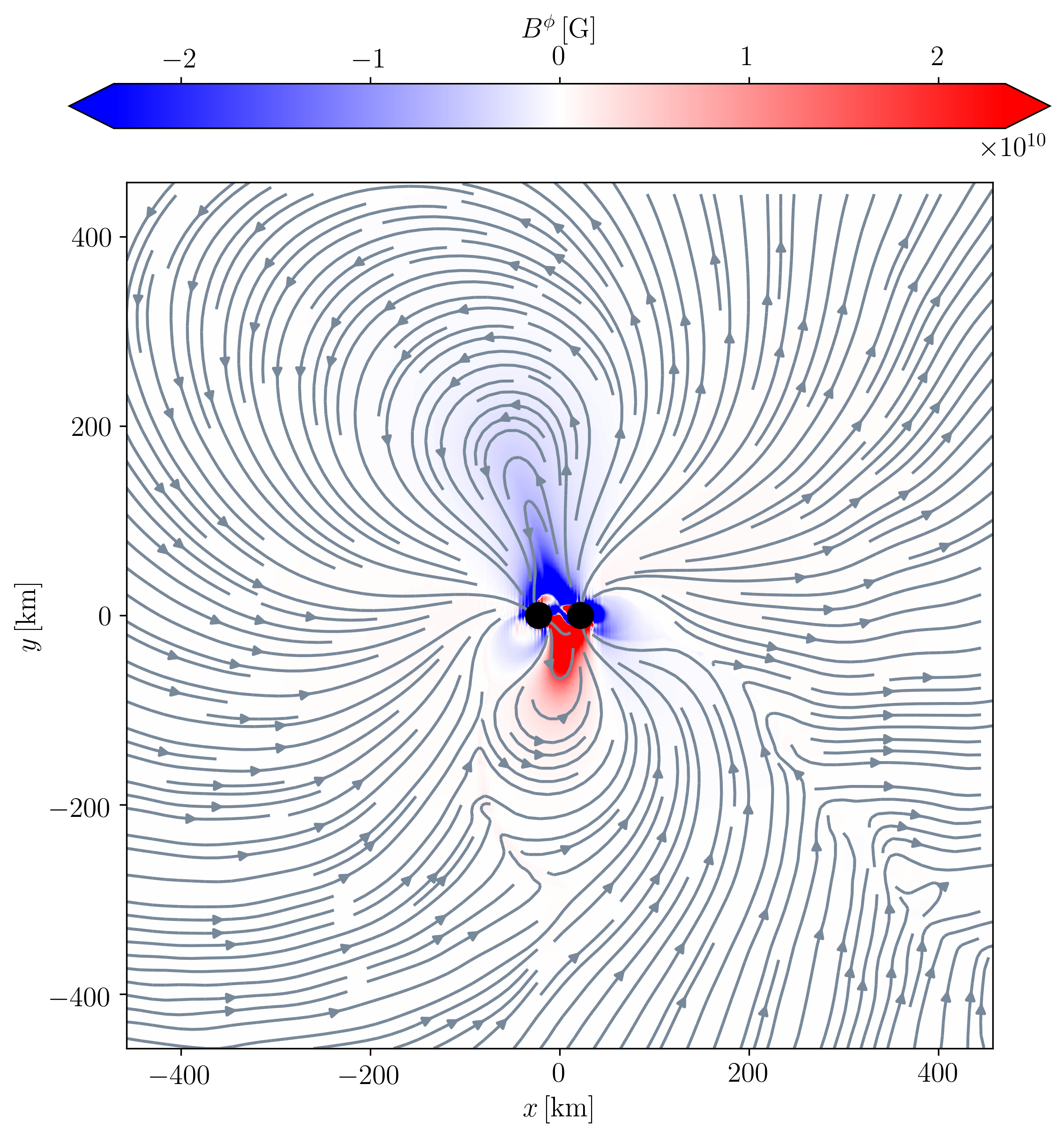}
  \hspace{0.1cm}
  \includegraphics[width=0.4\textwidth]{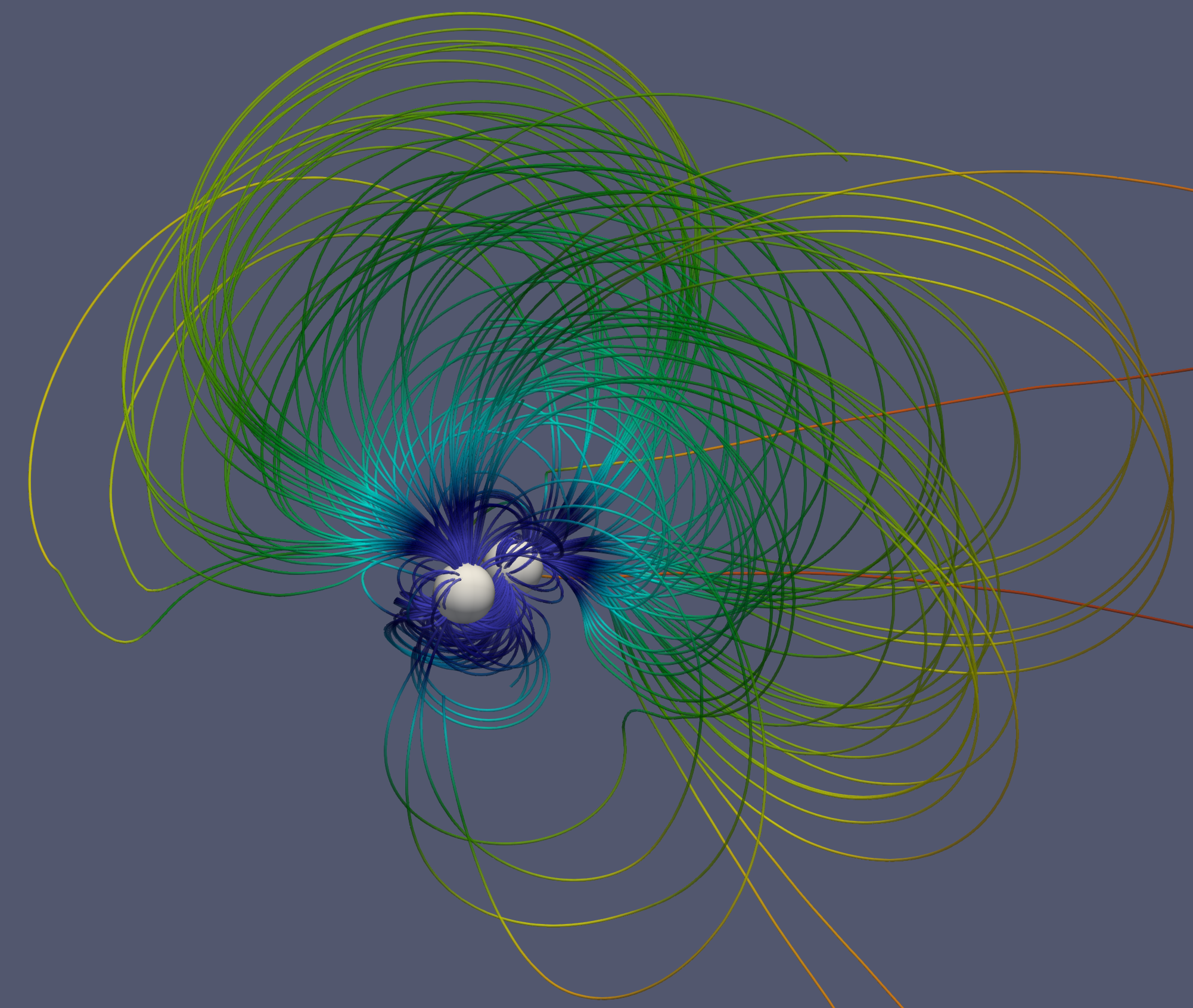}
  \caption{Some as in Fig. \ref{fig:emission_al} but with $45^\circ$ misaligned
   magnetic fields. The flaring is induced by the orbital motion. ({\it Top})
   Out-of-plane magnetic field $B^\phi$ in the meridional plane. 
   (Bottom) Three-dimensional visualisation of the field line configuration at the flare onset time.}
  \label{fig:emission2}
\end{figure}

Finally we highlight how the peak luminosity $\mathcal{L}_{\rm EM}^{\rm max}$  of the precursor flare for the fiducial binary scales with the orbital separation $a$. This is shown in Fig. \ref{fig:en_dep} for both the Poynting flux contained in the outgoing
magnetic bubble as well as for the dissipative luminosity in the current
sheet. We find a clean scaling of $\mathcal{L}_{\rm EM} \propto
a^{-7/2}$ in both cases.
In  order to better understand the scaling we estimate the {free} energy $\Delta E_{\rm twist}$ in the twisted flux tube in the limit of small twist, $\psi$, following \citet{Parfrey2013},
\begin{align}
  \Delta E_{\rm twist} &\approx \frac{1}{8} \psi^2 u^3 E_0 = \psi^2 \frac{R^3}{a^3} E_0,\\
  &\approx 1.4\times 10^{41} B^2_{12} R_{13}^3 \psi^2 \, {\rm erg},
\end{align}
where {$u$ characterizes the fraction of twisted field lines}, $u = 2 R /a$, and $E_0$ is the energy of the {unperturbed} dipole configuration.
In addition we have introduced $B_{12} = B/ 10^{12}\,\rm G$ and $R_{13} = R/13\,\rm km$.
If further {we assume that reconnection in the flaring sheet happens during the time} $\Delta t \simeq 2 a /{v_{\rm rec}}$ \citep{Parfrey2013}, where ${v_{\rm rec}} \sim 0.1 c$ is the reconnection speed, we find that
\begin{align}
  \mathcal{L}^{\max}_{\rm EM} = \frac{\Delta E_{\rm twist}}{\Delta t} &\approx \eta \frac{\psi^2}{2}v_{\rm rec}\frac{R^3}{a^4} E_0\\
  &\approx 4.6 \times 10^{44} \eta B^2_{12} R_{13}^3 \psi_\pi^2,
  \label{eqn:lum}
\end{align}
where {the pre-factor} $\eta\leq 1$ {quantifies the fraction of the {free} energy in the twisted flux tube that is available for reconnection, and flaring happens at twist value $\sim \pi$, i.e. $\psi_\pi = \psi/ \pi$}. Although not exact, this scaling is very similar to the one obtained from the simulations, 
and the differences might be associated with the assumption of small twist in Eq. \eqref{eqn:lum}.
We further find that the energy dissipated in the
current sheet is always and order of magnitude below the energy carried
away by the flare. Although the scaling found in our simulations seems very
clean we caution that we have not included any form of orbital decay 
caused by the inspiral. Since this will be subdominant at
larger separations we still believe our results to be applicable to most
binaries.

While our analysis has so far been focused on studying a single system
with equal magnetization and a non-zero stellar spin of one of the neutron stars we now
show that the flaring effect is quite general and will occur for a variety
of orbital configurations. In Fig. \ref{fig:emission2} we show the flaring process for an equally magnetized binary, where both stars are non-spinning but the left star has
a magnetic moment that is misaligned by $45^\circ$ with the direction of the orbital angular momentum. Whereas in the fiducial case the differential motion was caused by a relative difference in spin between the two stars. A non-spinning binary with a
misaligned magnetic field will twist its common magnetosphere because of the
orbital motion itself. In other words, while the energy that drives the flare in the
fiducial case was provided by the rotational motion of the star, it
is in this case provided purely by the orbital motion. Since we expect pulsar magnetic fields to be generally misaligned with their spin axis, this proves that magnetic flaring on the timescale of the orbital period is the most generic result of the interaction of magnetospheres with the comparable field strength of  both stars.

\section{Discussion}
\label{sec:conclusions}

We have presented the first force-free electrodynamics simulations
demonstrating how powerful electromagnetic flares can be launched as precursors
to the neutron star merger events. We have found that these are produced by a
built-up of twist in the common force-free magnetosphere of the binary system,
caused by differential motion. We have shown that this can be either caused by
a relative spin difference of the two stars or misalignment of the
magnetosphere, which are common for pulsars. 

While predicting both high-energy and coherent radio emission signatures of the
magnetic flare requires first-principles kinetic plasma simulations, lessons
learned in the pulsar magnetosphere research allow us to describe potential
outcomes. {For example, we expect most of the dissipated power in the current
  sheet to go into accelerated particles and, at typical magnetic field
  strengths $\geq 10^6$G at the sheet location, to be quickly radiated away as
  high-energy synchrotron radiation. Kinetic simulations of relativistic
  radiative reconnection (similar to ones done for pulsar current sheets in
  \citealt{Hakobyan2019}) that take into account synchrotron cooling of
  emitting particles and pair production due to collisions of high-energy
  photons are needed to calculate the expected high-energy signature. While
  this study adopts a simplified resistive force-free prescription for the
  dissipation in the current sheets, the $\approx 10\%$ efficiency of
  converting the outgoing Poynting flux into the dissipated power,
  $\mathcal{L}_{\rm dissipation}$, in magnetic reconnection is not uncommon in
  full kinetic plasma simulations of magnetospheres \citep{Philippov2014,
  Cerutti2015, Brambilla2018}. Future kinetic studies of flaring events in
  binary magnetospheres will be able to verify this conclusion
  \citep{Crinquand2019}.

Given relatively low expected luminosities in the high-energy band, coherent
radio emission is a best bet for a potential precursor signal
\cite{Lyutikov2019b}. We anticipate two potential mechanisms: the first channel
is a radio afterglow of the magnetic reconnection in the current sheet trailing
the outgoing bubble, and the second channel is the synchrotron maser emission
model of the outgoing magnetized bubble shocking the ambient plasma.
\cite{Lyubarsky2019} and \cite{Philippov2019} have shown that merging plasmoids
in the pulsar current sheet beyond the light cylinder can produce coherent
radio emission if the magnetic field strength in the upstream of the sheet is
$\simeq 10^6 - 10^8 \rm G$ \footnote{For higher field strength the sheet width
shrinks as $\propto B^{-3/2}$ \citep{Uzdensky2014b}, and the radiation will be
emitted at frequencies higher than radio.}. Re-scaling the results shown in
Figs. \ref{fig:emission_al} and \ref{fig:emission2} we find that in our
simulation this corresponds to fields strengths of $B_0 \simeq
10^{8}-10^{10}\rm G$ at the surface of the star. For such a field strength the
outgoing Poynting luminosity would, hence, be $\mathcal{L}_{\rm EM}^{\rm max}
\simeq 10^{39}-10^{41}\, {\rm erg s^{-1}}$. Given typically very low
efficiency, $\leq 10^{-4}$, of converting reconnecting magnetic flux into
escaping coherent electromagnetic waves it makes it unlikely to expect an
FRB-strength radio signal from the current sheet trailing the flare. As noted
by \cite{Lyubarsky2020} in a similar context of magnetar flares, reconnection
in the collision of the escaping magnetic bubble with the large-scale
magnetospheric current sheet is more likely to produce a powerful radio signal.
Another coherent radio emission channel is linked to the escaping magnetic
bubble. Since the ambient of the binary is not empty but filled with a highly
conducting electron-positron plasma with insignificant baryon pollution, the
bubble can produce a magnetized shock at some distance from the binary, which
may drive a synchrotron maser instability and associated electromagnetic
emission \citep{Gallant1992}. The quasi-isotropic structure of the escaping
magnetized bubble that we find in this work should help with the detectability
of these events. This scenario has been studied in the case of flares of young
magnetars, which is one possibility to explain Fast Radio Bursts
\citep{Lyubarsky2014, Beloborodov2017, Metzger2019}. We plan to explore these
possibilities in the future.}


One caveat of our study is that for now we have neglected effects of the
inspiral motion and general relativistic corrections, such as red shifts.
These will be particularly important to understand potential transients from the last orbits before the merger, and we plan to incorporate those effects in an upcoming work.
In order to better illustrate the observational prospect it will be important to cover
the vast parameter space of magnetic field configurations, e.g. to explore unequal magnetization and inclination effects. This investigation will be reported in a follow-up paper.

During the final preparation of this work we became aware of \cite{Carrasco2020}, who study the case of a single orbiting neutron star with a force-free magnetosphere using a setup very similar to the one described in this work. Their results focus on a continuous electromagnetic emission from a large-scale magnetospheric current sheet resulting from orbital motion, similar to the one that occurs in the magnetosphere of a usual rotating pulsar. While they are well applicable to systems with large orbital separations, i.e. greater than the stellar and orbital light cylinders, this work discusses the non-trivial magnetospheric interaction of two neutron stars prior to merger. It leads to the emission of powerful electromagnetic flares, which are more likely to be detectable as electromagnetic precursors of neutron star mergers \citep{Callister2019}.

\section*{Acknowledgements}
ERM and AP would like to thank Federico Carrasco, William East, Hayk Hakobyan, Luis Lehner, Yuri Levin, Maxim Lyutikov, Brian Metzger, Kohta Murase, Eliot Quataert, Luciano Rezzolla, Masaru Shibata, Lorenzo Sironi, Anatoly Spitkovsky and James Stone for valuable and insightful discussions. ERM gratefully acknowledges support and hospitality from the Simons Foundation through the pre-doctoral program at the Center for Computational Astrophysics, Flatiron Institute. This research was supported by the National Science Foundation under Grant No. AST-1909458. Research at the Flatiron Institute is supported by the Simons Foundation.

\bibliographystyle{yahapj}
\bibliography{aeireferences}

\begin{thebibliography}{}
\providecommand\natexlab[1]{#1}
\providecommand\JournalTitle[1]{#1}

\bibitem[{{Abbott} {et~al.}(2018){Abbott}, {Abbott}, {Abbott}, {Acernese},
  {Ackley}, {Adams}, {Adams}, {Addesso}, {Adhikari}, {Adya}, \&
  et~al.}]{Abbott2018b}
{Abbott}, B.~P., {Abbott}, R., {Abbott}, T.~D., {et~al.} 2018,
  \href{http://dx.doi.org/10.1103/PhysRevLett.121.161101}{\JournalTitle{Physical
  Review Letters}, 121, 161101}

\bibitem[{{Alic} {et~al.}(2012){Alic}, {Moesta}, {Rezzolla}, {Zanotti}, \&
  {Jaramillo}}]{Alic:2012}
{Alic}, D., {Moesta}, P., {Rezzolla}, L., {Zanotti}, O., \& {Jaramillo}, J.~L.
  2012,
  \href{http://dx.doi.org/10.1088/0004-637X/754/1/36}{\JournalTitle{Astrophys.
  J.}, 754, 36}

\bibitem[{{Bai} \& {Spitkovsky}(2010)}]{Bai2010}
{Bai}, X.-N., \& {Spitkovsky}, A. 2010,
  \href{http://dx.doi.org/10.1088/0004-637X/715/2/1282}{\JournalTitle{Astrophys.
  J.}, 715, 1282}

\bibitem[{{Baumgarte} \& {Shapiro}(2003)}]{Baumgarte2003}
{Baumgarte}, T.~W., \& {Shapiro}, S.~L. 2003,
  \href{http://dx.doi.org/10.1086/346103}{\JournalTitle{Astrophys. Journal},
  585, 921}

\bibitem[{{Beloborodov}(2017)}]{Beloborodov2017}
{Beloborodov}, A.~M. 2017,
  \href{http://dx.doi.org/10.3847/2041-8213/aa78f3}{\JournalTitle{Astrophys. J.
  Lett.}, 843, L26}

\bibitem[{{Bhattacharjee} {et~al.}(2009){Bhattacharjee}, {Huang}, {Yang}, \&
  {Rogers}}]{Bhattacharjee2009}
{Bhattacharjee}, A., {Huang}, Y.-M., {Yang}, H., \& {Rogers}, B. 2009,
  \href{http://dx.doi.org/10.1063/1.3264103}{\JournalTitle{Physics of Plasmas},
  16, 112102}

\bibitem[{Borges {et~al.}(2008)Borges, Carmona, Costa, \& Don}]{Borges2008}
Borges, R., Carmona, M., Costa, B., \& Don, W. 2008,
  \href{http://dx.doi.org/10.1016/j.jcp.2007.11.038}{\JournalTitle{Journal of
  Computational Physics}, 227, 3191}

\bibitem[{{Brambilla} {et~al.}(2018){Brambilla}, {Kalapotharakos}, {Timokhin},
  {Harding}, \& {Kazanas}}]{Brambilla2018}
{Brambilla}, G., {Kalapotharakos}, C., {Timokhin}, A.~N., {Harding}, A.~K., \&
  {Kazanas}, D. 2018,
  \href{http://dx.doi.org/10.3847/1538-4357/aab3e1}{\JournalTitle{Astrophys.
  J.}, 858, 81}

\bibitem[{{Callister} {et~al.}(2019){Callister}, {Anderson}, {Hallinan},
  {D'addario}, {Dowell}, {Kassim}, {Lazio}, {Price}, \&
  {Schinzel}}]{Callister2019}
{Callister}, T.~A., {Anderson}, M.~M., {Hallinan}, G., {et~al.} 2019,
  \href{http://dx.doi.org/10.3847/2041-8213/ab2248}{\JournalTitle{Astrophys. J.
  Lett.}, 877, L39}

\bibitem[{{Carrasco} \& {Shibata}(2020)}]{Carrasco2020}
{Carrasco}, F., \& {Shibata}, M. 2020, \JournalTitle{arXiv e-prints},
  arXiv:2001.04210

\bibitem[{{Carrasco} {et~al.}(2019){Carrasco}, {Vigan{\`o}}, {Palenzuela}, \&
  {Pons}}]{Carrasco2019}
{Carrasco}, F., {Vigan{\`o}}, D., {Palenzuela}, C., \& {Pons}, J.~A. 2019,
  \href{http://dx.doi.org/10.1093/mnrasl/slz016}{\JournalTitle{\mnras}, 484,
  L124}

\bibitem[{{Cerutti} {et~al.}(2015){Cerutti}, {Philippov}, {Parfrey}, \&
  {Spitkovsky}}]{Cerutti2015}
{Cerutti}, B., {Philippov}, A., {Parfrey}, K., \& {Spitkovsky}, A. 2015,
  \href{http://dx.doi.org/10.1093/mnras/stv042}{\JournalTitle{Mon. Not. R.
  Astron. Soc.}, 448, 606}

\bibitem[{{Crinquand} {et~al.}(2019){Crinquand}, {Cerutti}, \&
  {Dubus}}]{Crinquand2019}
{Crinquand}, B., {Cerutti}, B., \& {Dubus}, G. 2019,
  \href{http://dx.doi.org/10.1051/0004-6361/201834610}{\JournalTitle{Astron. \&
  Astrophys.}, 622, A161}

\bibitem[{{Crinquand} {et~al.}(2020){Crinquand}, {Cerutti}, {Philippov},
  {Parfrey}, \& {Dubus}}]{Crinquand2020}
{Crinquand}, B., {Cerutti}, B., {Philippov}, A.~e., {Parfrey}, K., \& {Dubus},
  G. 2020, \JournalTitle{arXiv e-prints}, arXiv:2003.03548

\bibitem[{{De} {et~al.}(2018){De}, {Finstad}, {Lattimer}, {Brown}, {Berger}, \&
  {Biwer}}]{De2018}
{De}, S., {Finstad}, D., {Lattimer}, J.~M., {et~al.} 2018,
  \href{http://dx.doi.org/10.1103/PhysRevLett.121.091102}{\JournalTitle{Physical
  Review Letters}, 121, 091102}

\bibitem[{Forbes(2000)}]{Forbes2000}
Forbes, T. 2000, \JournalTitle{Journal of Geophysical Research: Space Physics},
  105, 23153

\bibitem[{{Gallant} {et~al.}(1992){Gallant}, {Hoshino}, {Langdon}, {Arons}, \&
  {Max}}]{Gallant1992}
{Gallant}, Y.~A., {Hoshino}, M., {Langdon}, A.~B., {Arons}, J., \& {Max}, C.~E.
  1992, \href{http://dx.doi.org/10.1086/171326}{\JournalTitle{Astrophys. J.},
  391, 73}

\bibitem[{{Goldreich} \& {Julian}(1969)}]{Goldreich:1969}
{Goldreich}, P., \& {Julian}, W.~H. 1969,
  \href{http://dx.doi.org/10.1086/150119}{\JournalTitle{Astrophys. J.}, 157,
  869}

\bibitem[{{Hakobyan} {et~al.}(2019){Hakobyan}, {Philippov}, \&
  {Spitkovsky}}]{Hakobyan2019}
{Hakobyan}, H., {Philippov}, A., \& {Spitkovsky}, A. 2019,
  \href{http://dx.doi.org/10.3847/1538-4357/ab191b}{\JournalTitle{Astrophys.
  J.}, 877, 53}

\bibitem[{Hansen \& Lyutikov(2001)}]{Hansen2001}
Hansen, B. M.~S., \& Lyutikov, M. 2001,
  \href{http://dx.doi.org/10.1046/j.1365-8711.2001.04103.x}{\JournalTitle{Mon.
  Not. R. Astron. Soc.}, 322, 695}

\bibitem[{Kaspi \& Beloborodov(2017)}]{Kaspi2017}
Kaspi, V.~M., \& Beloborodov, A.~M. 2017,
  \href{http://dx.doi.org/10.1146/annurev-astro-081915-023329}{\JournalTitle{Annual
  Review of Astronomy and Astrophysics}, 55, 261}

\bibitem[{{Komissarov}(2004)}]{Komissarov2004b}
{Komissarov}, S.~S. 2004,
  \href{http://dx.doi.org/10.1111/j.1365-2966.2004.07598.x}{\JournalTitle{Mon.
  Not. R. Astron. Soc.}, 350, 427}

\bibitem[{{Lai}(2012)}]{Lai2012}
{Lai}, D. 2012,
  \href{http://dx.doi.org/10.1088/2041-8205/757/1/L3}{\JournalTitle{Astrophys.
  J. Lett.}, 757, L3}

\bibitem[{{Lehner} {et~al.}(2012){Lehner}, {Palenzuela}, {Liebling},
  {Thompson}, \& {Hanna}}]{Lehner2011}
{Lehner}, L., {Palenzuela}, C., {Liebling}, S.~L., {Thompson}, C., \& {Hanna},
  C. 2012,
  \href{http://dx.doi.org/10.1103/PhysRevD.86.104035}{\JournalTitle{Phys. Rev.
  D}, 86, 104035}

\bibitem[{{Loureiro} {et~al.}(2007){Loureiro}, {Schekochihin}, \&
  {Cowley}}]{Loureiro2007}
{Loureiro}, N.~F., {Schekochihin}, A.~A., \& {Cowley}, S.~C. 2007,
  \href{http://dx.doi.org/10.1063/1.2783986}{\JournalTitle{Physics of Plasmas},
  14, 100703}

\bibitem[{{Lyubarsky}(2014)}]{Lyubarsky2014}
{Lyubarsky}, Y. 2014,
  \href{http://dx.doi.org/10.1093/mnrasl/slu046}{\JournalTitle{Mon. Not. R.
  Astron. Soc.}, 442, L9}

\bibitem[{{Lyubarsky}(2019)}]{Lyubarsky2019}
---. 2019, \href{http://dx.doi.org/10.1093/mnras/sty3233}{\JournalTitle{Mon.
  Not. R. Astron. Soc.}, 483, 1731}

\bibitem[{{Lyubarsky}(2020)}]{Lyubarsky2020}
---. 2020, \JournalTitle{arXiv e-prints}, arXiv:2001.02007

\bibitem[{{Lyutikov}(2019{\natexlab{a}})}]{Lyutikov2019b}
{Lyutikov}, M. 2019{\natexlab{a}}, \JournalTitle{arXiv e-prints},
  arXiv:1901.03260

\bibitem[{{Lyutikov}(2019{\natexlab{b}})}]{Lyutikov2018}
---. 2019{\natexlab{b}},
  \href{http://dx.doi.org/10.1093/mnras/sty3303}{\JournalTitle{Mon. Not. R.
  Astron. Soc.}, 483, 2766}

\bibitem[{McCorquodale \& Colella(2011)}]{mccorquodale2011high}
McCorquodale, P., \& Colella, P. 2011, \JournalTitle{Communications in Applied
  Mathematics and Computational Science}, 6, 1

\bibitem[{{Metzger} {et~al.}(2019){Metzger}, {Margalit}, \&
  {Sironi}}]{Metzger2019}
{Metzger}, B.~D., {Margalit}, B., \& {Sironi}, L. 2019,
  \href{http://dx.doi.org/10.1093/mnras/stz700}{\JournalTitle{\mnras}, 485,
  4091}

\bibitem[{{Metzger} \& {Zivancev}(2016)}]{Metzger2016}
{Metzger}, B.~D., \& {Zivancev}, C. 2016,
  \href{http://dx.doi.org/10.1093/mnras/stw1800}{\JournalTitle{Mon. Not. R.
  Astron. Soc.}, 461, 4435}

\bibitem[{{Most} {et~al.}(2018{\natexlab{a}}){Most}, {Nathanail}, \&
  {Rezzolla}}]{Most2017}
{Most}, E.~R., {Nathanail}, A., \& {Rezzolla}, L. 2018{\natexlab{a}},
  \href{http://dx.doi.org/10.3847/1538-4357/aad6ef}{\JournalTitle{Astrophys.
  J}, 864, 117}

\bibitem[{{Most} {et~al.}(2019){Most}, {Papenfort}, \& {Rezzolla}}]{Most2019b}
{Most}, E.~R., {Papenfort}, L.~J., \& {Rezzolla}, L. 2019, \JournalTitle{arXiv
  e-prints}, arXiv:1907.10328

\bibitem[{{Most} {et~al.}(2020){Most}, {Papenfort}, \& {Rezzolla}}]{Most2020}
---. 2020, \JournalTitle{in preparation}, \href{http://arxiv.org/abs/in
  preparation}{{\sffamily arXiv:in preparation}}

\bibitem[{{Most} {et~al.}(2018{\natexlab{b}}){Most}, {Weih}, {Rezzolla}, \&
  {Schaffner-Bielich}}]{Most2018}
{Most}, E.~R., {Weih}, L.~R., {Rezzolla}, L., \& {Schaffner-Bielich}, J.
  2018{\natexlab{b}},
  \href{http://dx.doi.org/10.1103/PhysRevLett.120.261103}{\JournalTitle{Phys.
  Rev. Lett.}, 120, 261103}

\bibitem[{{Nathanail} {et~al.}(2017){Nathanail}, {Most}, \&
  {Rezzolla}}]{Nathanail2017}
{Nathanail}, A., {Most}, E.~R., \& {Rezzolla}, L. 2017,
  \href{http://dx.doi.org/10.1093/mnrasl/slx035}{\JournalTitle{Mon. Not. R.
  Astron. Soc.}, 469, L31}

\bibitem[{{Palenzuela}(2013)}]{Palenzuela2013}
{Palenzuela}, C. 2013,
  \href{http://dx.doi.org/10.1093/mnras/stt311}{\JournalTitle{Mon. Not. R.
  Astron. Soc.}, 431, 1853}

\bibitem[{{Palenzuela} {et~al.}(2010){Palenzuela}, {Lehner}, \&
  {Liebling}}]{Palenzuela:2010a}
{Palenzuela}, C., {Lehner}, L., \& {Liebling}, S.~L. 2010,
  \href{http://dx.doi.org/10.1126/science.1191766}{\JournalTitle{Science}, 329,
  927}

\bibitem[{{Palenzuela} {et~al.}(2013{\natexlab{a}}){Palenzuela}, {Lehner},
  {Liebling}, {Ponce}, {Anderson}, {Neilsen}, \& {Motl}}]{Palenzuela2013b}
{Palenzuela}, C., {Lehner}, L., {Liebling}, S.~L., {et~al.} 2013{\natexlab{a}},
  \href{http://dx.doi.org/10.1103/PhysRevD.88.043011}{\JournalTitle{Phys. Rev.
  D}, 88, 043011}

\bibitem[{{Palenzuela} {et~al.}(2013{\natexlab{b}}){Palenzuela}, {Lehner},
  {Ponce}, {Liebling}, {Anderson}, {Neilsen}, \& {Motl}}]{Palenzuela2013a}
{Palenzuela}, C., {Lehner}, L., {Ponce}, M., {et~al.} 2013{\natexlab{b}},
  \href{http://dx.doi.org/10.1103/PhysRevLett.111.061105}{\JournalTitle{Phys.
  Rev. Lett.}, 111, 061105}

\bibitem[{Pareschi \& Russo(2005)}]{pareschi_2005_ier}
Pareschi, L., \& Russo, G. 2005,
  \href{http://dx.doi.org/10.1007/BF02728986}{\JournalTitle{Journal of
  Scientific Computing}, 25, 129}

\bibitem[{{Parfrey} {et~al.}(2012){Parfrey}, {Beloborodov}, \&
  {Hui}}]{Parfrey2012}
{Parfrey}, K., {Beloborodov}, A.~M., \& {Hui}, L. 2012,
  \href{http://dx.doi.org/10.1088/2041-8205/754/1/L12}{\JournalTitle{Astrophys.
  J. Lett.}, 754, L12}

\bibitem[{{Parfrey} {et~al.}(2013){Parfrey}, {Beloborodov}, \&
  {Hui}}]{Parfrey2013}
---. 2013,
  \href{http://dx.doi.org/10.1088/0004-637X/774/2/92}{\JournalTitle{Astrophys.
  J.}, 774, 92}

\bibitem[{{Parfrey} {et~al.}(2019){Parfrey}, {Philippov}, \&
  {Cerutti}}]{Parfrey2019}
{Parfrey}, K., {Philippov}, A., \& {Cerutti}, B. 2019,
  \href{http://dx.doi.org/10.1103/PhysRevLett.122.035101}{\JournalTitle{Phys.
  Rev. Lett.}, 122, 035101}

\bibitem[{Paschalidis {et~al.}(2013)Paschalidis, Etienne, \&
  Shapiro}]{Paschalidis2013}
Paschalidis, V., Etienne, Z.~B., \& Shapiro, S.~L. 2013,
  \href{http://dx.doi.org/10.1103/PhysRevD.88.021504}{\JournalTitle{Phys. Rev.
  D}, 88, 021504}

\bibitem[{{Philippov} {et~al.}(2019){Philippov}, {Uzdensky}, {Spitkovsky}, \&
  {Cerutti}}]{Philippov2019}
{Philippov}, A., {Uzdensky}, D.~A., {Spitkovsky}, A., \& {Cerutti}, B. 2019,
  \href{http://dx.doi.org/10.3847/2041-8213/ab1590}{\JournalTitle{Astrophys. J.
  Lett.}, 876, L6}

\bibitem[{{Philippov} \& {Spitkovsky}(2014)}]{Philippov2014}
{Philippov}, A.~A., \& {Spitkovsky}, A. 2014,
  \href{http://dx.doi.org/10.1088/2041-8205/785/2/L33}{\JournalTitle{Astrophys.
  J.}, 785, L33}

\bibitem[{{Piro}(2012)}]{Piro2012}
{Piro}, A.~L. 2012,
  \href{http://dx.doi.org/10.1088/0004-637X/755/1/80}{\JournalTitle{Astrophys.
  J.}, 755, 80}

\bibitem[{{Ponce} {et~al.}(2014){Ponce}, {Palenzuela}, {Lehner}, \&
  {Liebling}}]{Ponce2014}
{Ponce}, M., {Palenzuela}, C., {Lehner}, L., \& {Liebling}, S.~L. 2014,
  \href{http://dx.doi.org/10.1103/PhysRevD.90.044007}{\JournalTitle{Phys. Rev.
  D}, 90, 044007}

\bibitem[{{Raithel}(2019)}]{Raithel2019}
{Raithel}, C.~A. 2019,
  \href{http://dx.doi.org/10.1140/epja/i2019-12759-5}{\JournalTitle{European
  Physical Journal A}, 55, 80}

\bibitem[{Rusanov(1961)}]{Rusanov1961a}
Rusanov, V.~V. 1961, \JournalTitle{J. Comput. Math. Phys. USSR}, 1, 267

\bibitem[{Schiff(1939)}]{Schiff39}
Schiff, L.~I. 1939,
  \href{http://dx.doi.org/10.1073/pnas.25.7.391}{\JournalTitle{Proceedings of
  the National Academy of Sciences}, 25, 391}

\bibitem[{the AMReX Development~Team {et~al.}(2019)the AMReX Development~Team,
  Almgren, Beckner, Blaschke, Chan, Day, Friesen, Gott, Graves, Katz, \&
  et~al.}]{Almgren2019}
the AMReX Development~Team, Almgren, A., Beckner, V., {et~al.}
  \href{http://dx.doi.org/10.5281/zenodo.2754709}{2019}

\bibitem[{{Uzdensky} \& {Spitkovsky}(2014)}]{Uzdensky2014b}
{Uzdensky}, D.~A., \& {Spitkovsky}, A. 2014,
  \href{http://dx.doi.org/10.1088/0004-637X/780/1/3}{\JournalTitle{Astrophys.
  J.}, 780, 3}

\bibitem[{{Wang} {et~al.}(2016){Wang}, {Yang}, {Wu}, {Dai}, \&
  {Wang}}]{Wang2016}
{Wang}, J.-S., {Yang}, Y.-P., {Wu}, X.-F., {Dai}, Z.-G., \& {Wang}, F.-Y. 2016,
  \href{http://dx.doi.org/10.3847/2041-8205/822/1/L7}{\JournalTitle{Astrophys.
  J. Lett.}, 822, L7}

\bibitem[{{Zhu} {et~al.}(2018){Zhu}, {Thrane}, {Os{\l}owski}, \&
  {Lasky}}]{ZhuX2018}
{Zhu}, X., {Thrane}, E., {Os{\l}owski}, Stefan~and{Levin}, Y., \& {Lasky},
  P.~D. 2018,
  \href{http://dx.doi.org/10.1103/PhysRevD.98.043002}{\JournalTitle{Phys. Rev.
  D}, 98, 043002}

\end{thebibliography}



\section*{Appendix}

In this appendix we briefly state the equations of general relativistic electrodynamics augmented with divergence cleaning in order to maintain the $D_i B^i=0$ and $D_i E^i= 4 \pi q$ constraints.
Written in covariant form using the field strength tensor $F^{\mu\nu}$ the augmented Maxwell equations read \citep{Palenzuela2013},
\begin{align}
\nabla_\mu \left(F^{\mu\nu} + \psi g^{\mu\nu}\right) 
&= - 4 \pi \mathcal{J}^\nu + \kappa_\psi \psi n^\nu, \label{eqn:Maxwell1_c}\\
  \nabla_\mu \left(\,^{\ast}F^{\mu\nu} + \phi g^{\mu\nu}\right) 
  &= \kappa_\phi \phi n^\nu \label{eqn:Maxwell2_c},
\end{align}
where $\phi,\psi$ are generalized Lagrange multipliers and $\kappa_\phi,\kappa_\psi$ are
their damping constants and the 4-current $\mathcal{J}^\mu = q n^\mu + J^\mu$,
where $J^\mu$ is the spatial part of the current, $q$ is the charge density and ${\bf n} = \alpha^{-1} \left( 1, - \beta^i \right)$ is the normal vector of the 3-dimensional hypersurface of the space-time foliation.

Within the 3+1 split of the metric introduced earlier in the text,
\begin{align}
{\rm d}s^2 = \left( -\alpha^2 + \beta_k \beta^k \right) {\rm d} t^2 
+ 2 \beta_k {\rm d} x^k {\mathrm d} t +\gamma_{ij} {\rm d}x^i {\rm d} x^j,    
\end{align}

the Maxwell equations \eqref{eqn:Maxwell1_c} and \eqref{eqn:Maxwell2_c} become \citep{Palenzuela2013},
\begin{align}
  \partial_t \left( \sqrt{\gamma} q \right) + \partial_i \left( \alpha J^i
  - \beta^i q \right) &=0, \label{eqn:qevol2}\\
  \partial_t \left( \sqrt{\gamma} B^i \right) + \partial_k \left(
  -\sqrt{\gamma}\beta^k B^i + \alpha \sqrt{\gamma} \varepsilon^{ikj} E_j  + \alpha \sqrt{\gamma}
  \phi\right)
  &= -\sqrt{\gamma} B^k \partial_k \beta^i + \sqrt{\gamma} \phi \left(
  \gamma^{ij} \partial_j \alpha - \alpha \gamma^{jk} \Gamma^i_{jk}
  \right),\\
  \partial_t \left( \sqrt{\gamma} \phi \right) + \partial_k \left( - \beta^k
    \sqrt{\gamma}\phi + \alpha \sqrt{\gamma} B^k
    \right)
    &=  - \alpha \sqrt{\gamma} \phi K +
    \sqrt{\gamma} B^k \partial_k \alpha - \alpha \kappa_\phi \sqrt{\gamma}
    \phi, \\
  \partial_t \left( \sqrt{\gamma}E^i \right) + \partial_k \left(
  -\sqrt{\gamma}\beta^k E^i - \alpha \sqrt{\gamma} \varepsilon^{ikj} B_j  + \alpha \sqrt{\gamma}
  \psi\right)
  &= -\sqrt{\gamma} E^k \partial_k \beta^i + \sqrt{\gamma} \psi \left(
  \gamma^{ij} \partial_j \alpha - \alpha \gamma^{jk} \Gamma^i_{jk}
  \right) - 4 \pi \alpha \sqrt{\gamma} J^i,\\
  \partial_t \left( \sqrt{\gamma} \psi \right) + \partial_k \left( - \beta^k
    \sqrt{\gamma}\psi + \alpha \sqrt{\gamma} E^k
    \right)
    &=  - \alpha \sqrt{\gamma} \psi K +
    4 \pi \alpha \sqrt{\gamma} q+
    \sqrt{\gamma} E^k \partial_k \alpha - \alpha \kappa_\psi \sqrt{\gamma} \psi.\label{eqn:Bevol}
\end{align}
Here $\Gamma^k_{ij}$ is the metric compatible Christoffel symbol of the 3-metric $\gamma_{ij}$, $\gamma = \det \gamma_{ij}$ and $K$ is the trace of the extrinsic curvature.
These equations are covariant and are valid in any frame, in particular also in the corotating frame \citep{Schiff39}.
In order to clarify the implications of this approach and similar (but not identically) to \cite{Carrasco2020},
we adopt a co-rotating Minkowski
frame \citep{Schiff39} $\alpha=1$, $\gamma_{ij} = \delta_{ij}$ and $\beta_i = -
\varepsilon_{ijk} \Omega^j x^k$, where ${\bf \Omega} = (0,0,0,\omega)$ is the
orbital angular momentum vector. In order to understand the meaning of this construction we consider the advection velocity $u^\mu$ of a fluid element inside the stars.\\
There,
\begin{align}
        \frac{u^i}{u^0} = \alpha v^i - \beta^i = \varepsilon^{ijk} \Omega_{s\, j} x_{s\, k} + \varepsilon^{ijk} \Omega_j x_k,
\end{align}
where $v^i$ is the 3-velocity seen by the corotating observer. In simple words, since the observer corotates with the star, the local velocity he sees does not contain the $\bf \Omega \times x$ part of the orbital motion and, hence, does not enter the electric field
\begin{align}
  E^i = - \varepsilon^{ijk} v_j B_k,
  \label{eqn:E}
\end{align}
seen by the corotating observer. Nonetheless, because we are solving the covariant form of the Maxwell equations, we can see that the actual field that enters the induction equation is $\boldsymbol E - \boldsymbol \beta \times \boldsymbol B$, and hence the rotational contribution does enter into the magnetic field evolution consistently.
Since $\bf \Omega$ and $\bf \Omega_s$ are aligned we can express this as
\begin{align}
      \frac{u^i}{u^0} =  \varepsilon^{ijk} \left(\Omega_s + \Omega\right) z_j x_{s\, k} + \varepsilon^{ijk} \Omega_j x_{\rm star \,k},
\end{align}
where $\bf z$ is the z-coordinate vector and ${\bf x_{\rm star}} = {\bf x} - \bf x_s$ is the coordinate vector of the stellar center hence.
Hence, a fluid element inside the star would spin with $\left(\Omega_s + \Omega\right)$ around the spin axis of the star and at the same time co-move with the orbital motion of the star.
If we would in addition subtract $\varepsilon_{ijk} \Omega^j x_{\rm star}^k$ from the 3-velocity $v^i$, which enters the electric field inside the star via Eq. \eqref{eqn:E},
we would indeed adopt a truly corotating frame in which the star no longer moves since then
\begin{align}
    \frac{u^i}{u^0} = \alpha v^i - \beta^i = \varepsilon^{ijk} \left(\Omega_s + \Omega\right) z_j x_{s\, k},
\end{align}
which corresponds to the purely rotational motion of the star.
We can also see from this expression that $\Omega_s=0$ implies a residual rotation in the true comoving frame, corresponding to the case of synchronised orbital motion of the stars.

\end{document}